# Identifying and Understanding the Positive Impact of Defects for Optoelectronic Devices

Yu-Hao Deng[1*]

[1] Physics and Chemistry of Nanostructures Group, Ghent University, 9000 Ghent, Belgium

* Correspondence should be addressed to yuhao.deng@ugent.be

**Abstract**

Defects are generally regarded to have negative impacts on carrier recombination, charge-transport and ion migration in materials, which thus lower the efficiency, speed and stability of optoelectronic devices. Meanwhile, lots of efforts which focused on minimizing defects have greatly improved the performances of devices. Then, can defects be positive in optoelectronic devices? Herein, relying on in-depth understanding of defect-associated effects in semiconductors, trapping of photo-generated carriers by defects is applied to enlarge photoconductive gain in photodetection. Therefore, the record photoconductive gain, gain-bandwidth product and detection limit were achieved in this photodetector. Exceeding the general concept that defects are harmful, we identify a new view that the defects can be positive in photodetection, which may guide us to design high-performance photodetectors.

## Introduction

Optoelectronic devices, such as lasers, and light-emitting diodes (LEDs), image sensors, solar cells and so on, changed our life-style completely in the last decades [1, 2]. Despite the rapid progress of optoelectronics, defects are usually recognized as the key factor to limit the improvement of device performance [3-5]. As shown in Fig. 1A, defects that cause electronic states within the semiconductor bandgap would capture or trap the approaching electrons or holes. The trapped carriers will annihilate or recombine with opposite carriers and induce non-radiative recombination, then accompanied by the emission of phonons. This pathway of recombination for the photo-generated carriers is undesirable and is considered an important loss mechanism in solar cells [6, 7]. Fig. 1B indicates that carrier transport is also affected by defects. Defects in semiconductors break the periodicity of the crystal structure and/or form charge centers, resulting in deflected scattering of free carriers [8]. This detrimental scattering process lowers the charge carrier mobility (μ) of semiconductors. Moreover, as shown in Fig. 1C, the induced defects by vacancies and interstitials, introduce the ion migration channels in semiconductor films, have been demonstrated to play a vital role in the chemical degradation of perovskite materials [9, 10]. Thus far, defects are generally regarded as harmful to optoelectronic devices and most of efforts are focused on suppression effects in materials. So, it remains a question that can defects be positive in optoelectronic devices?

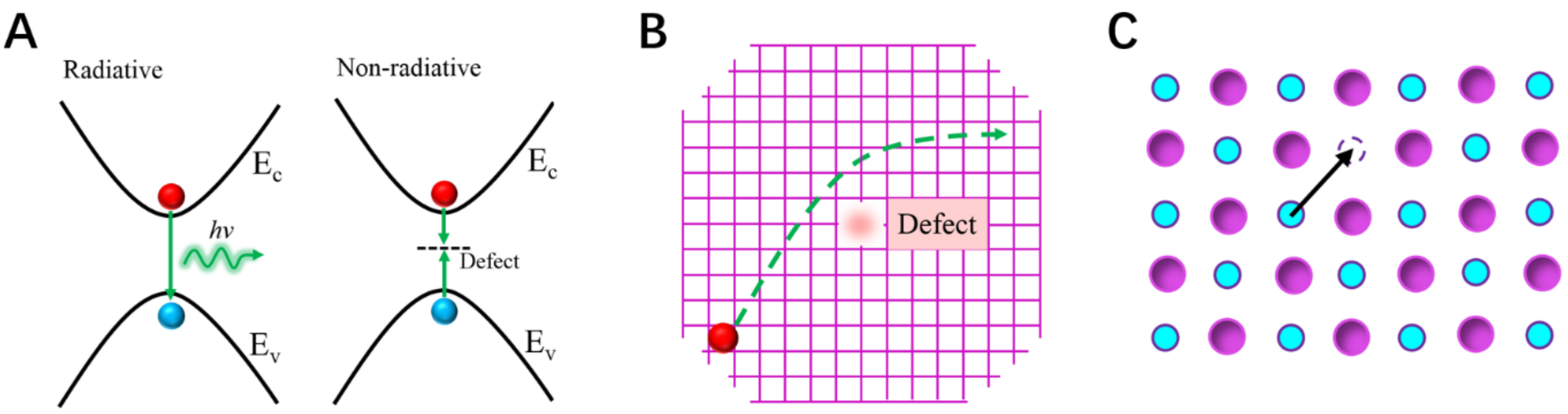

**Fig. 1. Effects associated with defects in semiconductors. (A)** Radiative recombination pathway of photo-generated carriers and electron and hole are captured by defect in non-radiative recombination. **(B)** Scattering of carrier near ionized point defect. **(C)** Ionic transport involving conventional vacancy hopping between neighboring positions.

## Photoconductive gain in photodetector

The photoconductive gain ($G$) is the ability of optoelectronic devices to amplify photons, thus high-gain in photodetector is essential for highly sensitive sensors and electronic amplifying circuits [11, 12]. The $G$ is defined as the number of collected carriers' worth of absorbed photons per unit time, which is equivalent to the ratio of photon-generated carrier lifetime and the carrier transit time [13]. Transit time equals to $d^2/\mu V$, where $d$ is the thickness of active layer and $V$ is the applied external voltage [12]. Gain in optoelectronic devices can be expressed as

$$G = \frac{N_e}{N_p} = \frac{\tau}{\tau_{tr}}\eta = \frac{\tau\mu V}{d^2}\eta \qquad (1)$$

where $N_e$ and $N_e$ are the number of collected carriers and absorbed photons, $\tau$ is the lifetime of photon-generated carrier, $\tau_{tr}$ is the carrier transit time, $\eta$ is the external quantum efficiency of the device [12-16].

## Gain acquisition via carrier trapping

Lack of stoichiometric compositions at the surfaces of semiconductor films form defects (Fig. 2A). The surface defects can be used to trap photo-generated carriers, which prevent charge-carrier recombination and prolongs the carrier lifetime ($\tau$). As the schematic of photodetector shown in Fig. 2B, upon illumination with photon energy above semiconductor bandgap, electron-hole pairs are generated and electrons are readily trapped at the semiconductor/metal interface. The lifetime of trapped electrons is much longer than the transit time of free holes. Under an applied external voltage, the unpaired holes are collected at the cathode. When the electrons are trapped, the number of holes injected in the active layers will pass through the circuit before recombining with electrons, which greatly exceeds the number of the originally photo-generated carriers. This gain theory is also called as "recycling gain mechanism" [17].

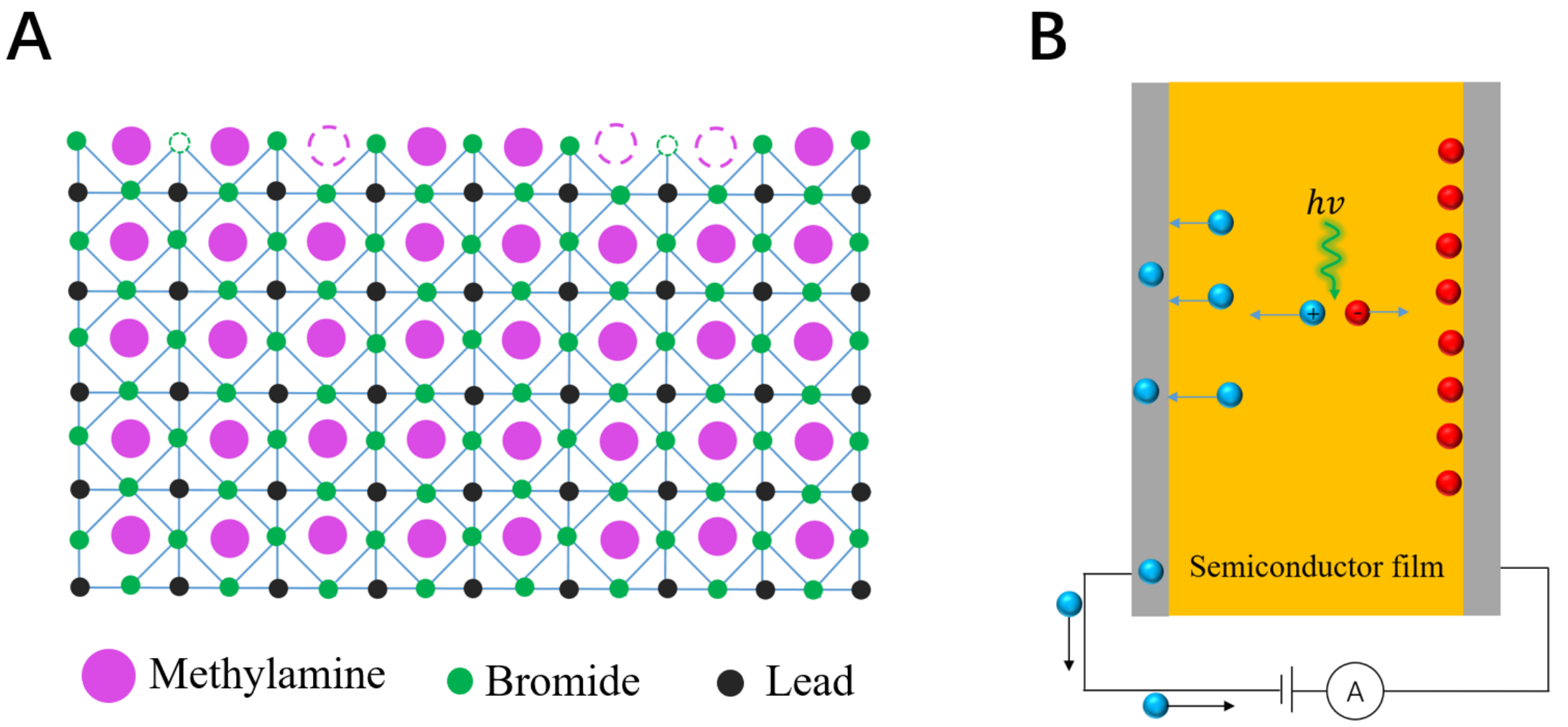

**Fig. 2. Surface defects and trapping mechanism in perovskite film. (A)** Illustration of point defects in perovskites. Here, colors represent the following: black, lead; green, bromide; purple, methylamine. **(B)** Schematic of gain mechanism in photodetector.

## Performance improvement by device design

Based on equation (1), prolong the carrier lifetime ($\tau$) and reduce carrier transit time ($\tau_{tr}$) can enlarge $G$. Meanwhile, improving the quality of semiconductor and ensuring the full absorption of light indicate a more efficient $\eta$. Furthermore, in order to reduce the carrier transit time, the thickness of active layer should be optimized to be as thin as possible while ensuring full absorption of incident light. The carrier mobility ($\mu$) of the active layer also should be improved as high as possible to increase the carrier transport velocity and reduce the $\tau_{tr}$, so it is preferable to employ a single-crystal material as active layer rather than polycrystalline. Moreover, applying the purification for the precursor to reduce internal defects in single-crystal perovskite film, the photodetector achieved a record photoconductive gain of 50 million and a gain-bandwidth product of 70 GHz. The gain and the gain-bandwidth product of the device were over 5000 times and two orders of magnitude higher than all the reported perovskite photodetectors based on photoconductive mechanism respectively (Fig.3A) [14]. The device also obtained an ultrahigh sensitivity with record detection limit down to about 200 photons, which is over 50 times lower than all reported perovskite photodetectors (Fig.3B) [14].

In this example, we used the defects on the surface of perovskite layer to trap the carrier and prolong the $\tau$. High-mobility single crystalline perovskite layer was used to maximize carrier transport velocity and decrease the $\tau_{tr}$. Moreover, the thickness of active layer was optimized to be as thin as possible to decrease the $\tau_{tr}$ while ensuring full absorption of incident light. Due to these optimizations, the record performances were achieved in our photodetector.

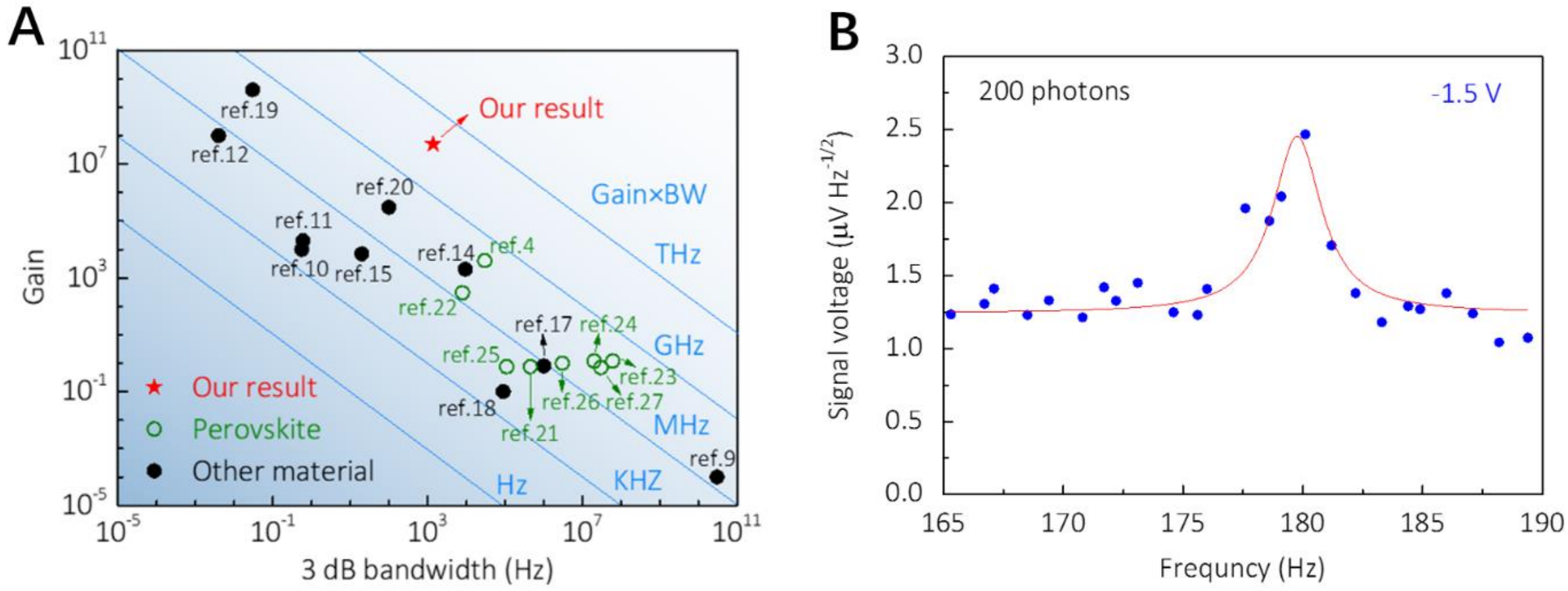

**Fig. 3. Record performances of the photodetector. (A)** Gain and Gain-bandwidth product of the detectors based on different materials. **(B)** The device responsed to 200 photons in each pulse of the incident light at 180 Hz at -1.5 V bias. (A, B) Reproduced with permission from Ref. [14], ©WILEY-VCH Verlag GmbH & Co. KGaA, Weinheim 2018.

## Conclusion and outlook

Exceeding the general concept that defects are harmful, we identify that the defects can be positive in optoelectronic devices to improve the sensitivity. With the help of surface defects, the record photoconductive gain, gain-bandwidth product and detection limit were achieved in our perovskite photodetector. The trapping of photo-carriers by defects decreases the performance of photovoltaic performance devices but improves the sensitivity of photodetectors. The trap states not only can be applied to trap carriers on the surface of active layer, but also in the active layer and on the surface of transport layers. Dong et al. mixed ZnO nanoparticles with PbS QDs to leave long-lifetime trapped electrons on ZnO particles, got a gain as 16 in this hybrid photodetector [18]. Xu et al. introduced the oxygen vacancy defects at the ZnO surface to trap electrons in electron transport layer, a gain as 153 was reported in PbS QDs photodetector [19]. High gain is obtained by prolonging carrier lifetime, which means it also increases the response time of photodetector. Therefore, achieving ultrafast and gain-based high-sensitivity devices simultaneously will represent a significant challenge. In addition to increasing the gain in the photodetector, defects could also be positively used in emitter [20], even laser, and quenching in quantum dot spectrometers [21]. In general, in-depth understanding of defects evolution and delicate design are the keys to maximize performances of optoelectronic devices. This new view and understanding may guide us to design highly sensitive photodetectors in the future.

**Data availability**

All data are available from the corresponding author(s) upon reasonable request.

**Conflict of interest:** The authors declare no competing financial interest.

**Note:** This manuscript has been published in *Advanced Sensor Research* [22].

**References**

1. Bhattacharya, Pallab. Semiconductor optoelectronic devices. Prentice-Hall, Inc., 1997.
2. Brennan, Kevin F. The physics of semiconductors: with applications to optoelectronic devices. Cambridge university press, 1999.

3. Reshchikov, Michael A., and Hadis Morkoç. "Luminescence properties of defects in GaN." Journal of applied physics 97.6 (2005).
4. Xu, Yong, et al. "Doping: a key enabler for organic transistors." Advanced Materials 30.46 (2018): 1801830.
5. Chen, Bo, et al. "Imperfections and their passivation in halide perovskite solar cells." Chemical Society Reviews 48.14 (2019): 3842-3867.
6. Hall R N. Electron-hole recombination in germanium[J]. Physical review, 1952, 87(2): 387.
7. Shockley W, Read Jr W T. Statistics of the recombinations of holes and electrons[J]. Physical review, 1952, 87(5): 835.
8. Conwell E, Weisskopf V F. Theory of impurity scattering in semiconductors[J]. Physical review, 1950, 77(3): 388.
9. Eames C, Frost J M, Barnes P R F, et al. Ionic transport in hybrid lead iodide perovskite solar cells[J]. Nature communications, 2015, 6(1): 1-8.
10. Yoon S J, Kuno M, Kamat P V. Shift happens. How halide ion defects influence photoinduced segregation in mixed halide perovskites[J]. ACS Energy Letters, 2017, 2(7): 1507-1514.
11. Konstantatos G, Clifford J, Levina L, et al. Sensitive solution-processed visible-wavelength photodetectors[J]. Nature photonics, 2007, 1(9): 531-534.
12. Soci C, Zhang A, Xiang B, et al. ZnO nanowire UV photodetectors with high internal gain[J]. Nano letters, 2007, 7(4): 1003-1009.
13. Saleh B E A, Teich M C. Fundamentals of photonics[M]. john Wiley & sons, 2019.
14. Yang Z, Deng Y, Zhang X, et al. High-Performance Single-Crystalline Perovskite Thin-Film Photodetector[J]. Advanced Materials, 2018, 30(8): 1704333.
15. J. M. Liu, Photonic Devices, Cambridge University Press, New York 2009.
16. Deng, Yu-Hao, Zhen-Qian Yang, and Ren-Min Ma. "Growth of centimeter-scale perovskite single-crystalline thin film via surface engineering." Nano convergence 7 (2020): 1-7.
17. Neamen, Donald. Semiconductor physics and devices. McGraw-Hill, Inc., 2002.
18. Dong, Rui, et al. "An ultraviolet-to-NIR broad spectral nanocomposite photodetector with gain." Advanced Optical Materials 2.6 (2014): 549-554.

19. Xu, Kaimin, et al. "Large photomultiplication by charge-self-trapping for high-response quantum dot infrared photodetectors." ACS Applied Materials & Interfaces 14.12 (2022): 14783-14790.
20. Salh, Roushdey. "Defect related luminescence in silicon dioxide network: a review." Crystalline Silicon-Properties and Uses (2011): 135-172.
21. Bao, Jie, and Moungi G. Bawendi. "A colloidal quantum dot spectrometer." Nature 523.7558 (2015): 67-70.
22. Deng, Yu-Hao. Identifying and Understanding the Positive Impact of Defects for Optoelectronic Devices. Advanced Sensor Research (2024), 2300144. https://doi.org/10.1002/adsr.202300144